\documentclass[aps,prd,amsmath,preprint]{revtex4}

\usepackage{amsmath}
\usepackage{bm}
\usepackage[dvips]{color,graphicx}
\usepackage{epsfig}
\usepackage{dcolumn}

\newcommand{\eps}{\varepsilon}
\newcommand{\sw}{\sin^2\theta_W}
\newcommand{\cw}{\cos^2\theta_W}

\begin{document}

\title{Effect of two-boson exchange on	% \\
	parity-violating $ep$ scattering}

\author{J. A. Tjon}
\affiliation{Physics Department, University of Utrecht,
		The Netherlands}

\author{W. Melnitchouk}
\affiliation{Jefferson Lab, 12000 Jefferson Avenue,
		Newport News, VA 23606, USA}

\begin{abstract}
We compute the corrections from two-photon and $\gamma$--$Z$
exchange in parity-violating elastic electron--proton scattering,
used to extract the strange form factors of the proton.
We use a hadronic formalism that successfully reconciled the earlier
discrepancy in the proton's electron to magnetic form factor ratio,
suitably extended to the weak sector.
Implementing realistic electroweak form factors, we find effects
of the order 2--3\% at $Q^2 \lesssim 0.1$~GeV$^2$, which are largest
at backward angles, and have a strong $Q^2$ dependence at low $Q^2$.
Two-boson contributions to the weak axial current are found to be
enhanced at low $Q^2$ and for forward angles.
We provide corrections at kinematics relevant for recent and
upcoming parity-violating experiments.
\end{abstract}

\maketitle

%%%%%%%%%%%%%%%%%%%%%%%%%%%%%%%%%%%%%%%%%%%%%%%%%%%%%%%%%%%%%%%%%%%%%%%%%
Two-photon exchange corrections have recently been found to play an
unexpectedly important role in elastic electron--proton scattering.
Despite being of ${\cal O}(\alpha)$ smaller than the Born amplitudes,
$2\gamma$ exchange effects have been shown to have a strong angular
dependence, which significantly influences the extraction of the
electric form factor in Rosenbluth separations.
Such corrections were found to resolve a major part of the discrepancy
between the electric to magnetic proton form factor ratio,
$G_E^p/G_M^p$, determined via the Rosenbluth and polarization
transfer methods (see Ref.~\cite{BMT} and references therein).

Elastic $ep$ scattering has also been used to probe the strangeness
content of the proton, through measurements of the strange electric
and magnetic form factors.
This is achieved by scattering polarized electrons from unpolarized
protons, and observing the parity-violating (PV) asymmetry
$A_{\rm PV} = (\sigma_R - \sigma_L)/(\sigma_R + \sigma_L)$,
where $\sigma_{R(L)}$ is the cross section for a right- (left-)
handed electron.
The numerator in the asymmetry is sensitive to the interference of
the vector and axial-vector currents, and hence violates parity
\cite{Musolf}.

In view of the large $2\gamma$ contributions to electromagnetic form
factors, it is natural to ask what effect the exchange of two bosons
($\gamma$ or $Z$) may have on the PV asymmetries.
In particular, since the extracted strange form factors appear to be
rather small \cite{Ross}, these two-boson exchange (TBE) contributions
could affect the extraction significantly.
In this paper we compute the relevant corrections to the PV asymmetry
arising from the interference between single $Z$ boson and $2\gamma$
exchange amplitudes (which we denote by ``$Z(\gamma\gamma)$''),
and between the one-photon exchange and $\gamma$--$Z$ interference
amplitudes (denoted by ``$\gamma(Z\gamma)$'').
We use the hadronic formalism developed in Ref.~\cite{BMT}, which
allows a more natural implementation of hadronic structure effects
in radiative corrections at low four-momentum transfer squared $Q^2$,
where PV electron scattering experiment are typically performed.

In their seminal work on electroweak radiative effects, Marciano \&
Sirlin \cite{Marciano} computed the $\gamma(Z\gamma)$ contribution
at $Q^2=0$, both at the quark level and at the hadronic level using
dipole form factors.
The $Z(\gamma\gamma)$ contribution was calculated in Ref.~\cite{AC}
within a generalized parton distribution formalism, applicable at a
scale of several GeV$^2$.
More recently, Zhou {\em et al.} \cite{Yang} computed the TBE effects on
$A_{\rm PV}$ within a hadronic basis using monopole form factors.

In the present analysis, we account for the finite size of the proton
by using realistic electromagnetic form factors in the loop graphs,
determined self-consistently from a global analysis of elastic cross
section and polarization transfer data \cite{AMT} including explicit
$2\gamma$ exchange corrections.
Furthermore, while an overall, factorized correction was applied
to the PV asymmetry in Ref.~\cite{Yang}, here we compute explicitly
the individual TBE corrections to the proton and neutron (or to the
$\sw$-dependent and independent) terms in $A_{\rm PV}$.

In the Born approximation, the amplitude for the weak current mediated
by $Z$ exchange is given by:
\begin{equation}
{\cal M}_Z
= \frac{e^2}{Q^2+M_Z^2}\
   \frac{1}{(4 \sw \cw)^2}\
   j_Z^\mu\ J_{Z \mu}\ ,
\end{equation}
where $\sw = 1 - M_W^2/M_Z^2$ is the Weinberg angle,
with $M_W$ ($M_Z$) the $W$ ($Z$) boson mass.
For the corresponding electromagnetic one-photon exchange amplitude,
${\cal M}_\gamma$, we refer to Ref.~\cite{BMT} for details.
The weak leptonic current is given by a sum of vector and
axial-vector terms, 
$j_Z^\mu
= \bar{u}_e (g^e_V \gamma^\mu + g^e_A \gamma^\mu \gamma_5) u_e$,
where $g^e_V = -(1-4 \sin^2\theta_W)$ and $g^e_A=+1$ are the vector
and axial-vector couplings of the electron to the weak current.
The matrix element of the weak nucleonic matrix is given by
$J_Z^\mu = \bar{u}_N \Gamma_Z^\mu u_N$, where the current operator
is parameterized by the weak form factors:
\begin{equation}
\Gamma_Z^\mu
= \gamma^\mu\ F_1^Z
+ \frac{i \sigma^{\mu\nu} q_\nu}{2 M}\ F_2^Z
+ \gamma^\mu \gamma_5\ G_A^Z\ ,
\label{eq:JZ}
\end{equation}
where $q$ is the four-momentum transfer and $M$ is the nucleon mass.  
Assuming isospin symmetry, the weak vector form factors $F_{1,2}^{Z p}$
for a proton target are related to the electromagnetic form factors of
the proton and neutron $F_{1,2}^{\gamma p,n}$ (at tree level) by
\cite{Musolf}:
\begin{equation}
\label{eq:FZ}
F_{1,2}^{Z p}
= (1-4 \sw) F_{1,2}^{\gamma p} - F_{1,2}^{\gamma n} - F_{1,2}^s\ ,
\end{equation}
where $F_{1,2}^s$ are the contributions from strange quarks.
The small factor $(1-4 \sw)$ suppresses the overall contribution
from the proton electromagnetic form factors.
The weak axial-vector form factor of the proton is given by
$G_A^{Z p} = -G_A^{\gamma p} + G_A^s$, where $G_A^s$ is the
strange quark contribution.

In the standard model the parity-violating asymmetry $A_{\rm PV}$
receives contributions from products of vector electron and
axial-vector proton currents, and axial-vector electron and vector
proton currents.
It can be written as a sum of proton vector, strange, and
axial-vector contributions:
\begin{eqnarray}
A_{\rm PV}
= - \left( {G_F Q^2 \over 4\sqrt{2} \pi \alpha} \right)
    \left( A_V + A_s + A_A \right)\ ,
\end{eqnarray}
where 
\begin{subequations}
\label{eq:A}
\begin{align}
\label{eq:AV}
A_V
&= g^e_A\ \rho
\left[
  (1-4\kappa\sw)
- { \left( \eps G_E^{\gamma p} G_E^{\gamma n}
	 + \tau G_M^{\gamma p} G_M^{\gamma n}
    \right) \over \sigma_{\rm red} }
\right],				\\
\label{eq:As}
A_s
&= - g^e_A\ \rho\
{ \left( \eps G_E^{\gamma p} G_E^s
       + \tau G_M^{\gamma p} G_M^s
  \right) \over \sigma_{\rm red} }\ ,		\\
\label{eq:AA}
A_A
&= g^e_V\ \sqrt{\tau (1+\tau)(1-\eps^2)}\
{ \widetilde{G}_A^{Z p} G_M^{\gamma p}
  \over \sigma_{\rm red} }\ ,
\end{align}
\end{subequations}
where $G_F = \pi \alpha/(\sqrt{2} \sw M_W^2)$ is the Fermi constant,
and
$\sigma_{\rm red} = \eps (G_E^{\gamma p})^2 + \tau (G_M^{\gamma p})^2$
is the reduced unpolarized cross section,
with $\varepsilon = (1 + 2(1+\tau)\tan^2(\theta/2))^{-1}$ the photon
polarization parameter and $\tau = Q^2/4M^2$.

The parameters $\rho$ and $\kappa$ in Eqs.~(\ref{eq:A}) contain higher
order radiative effects, such as vertex corrections, wave function
renormalization, vacuum polarization, and inelastic bremsstrahlung,
which have been calculated previously and are well known.
At tree level, $\rho = \kappa = 1$.
Beyond tree level, $\rho$ and $\kappa$ also contain contributions
from the interference of the Born and TBE diagrams, which we denote
by $\Delta\rho$ and $\Delta\kappa$, respectively.
The form factor $\widetilde{G}_A^{Z p}$ implicitly contains higher
order radiative corrections for the proton axial current, as well as
the hadronic anapole contributions \cite{Musolf,Ross}.
At tree level, and in the absence of the anapole term,
$\widetilde{G}_A^{Z p} \to G_A^{Z p}$ above.

The contribution of the $Z(\gamma\gamma)$ and $\gamma(Z\gamma)$
TBE corrections to the PV cross section can be written:
\begin{eqnarray}
\Delta\sigma^{\rm TBE}
= 2\ \Re
\left[ {\cal M}_{\gamma\gamma} {\cal M}_Z^*
     + ({\cal M}_{\gamma Z} + {\cal M}_{Z\gamma}) {\cal M}_\gamma^*
\right],
\end{eqnarray}
where ${\cal M}_{\gamma\gamma}$ (${\cal M}_{\gamma Z}$)
is the two-photon ($\gamma$--$Z$) exchange amplitude.
Since the asymmetry $A_{\rm PV}$ is constructed as a ratio of
the PV cross section to the unpolarized (parity conserving)
cross section, one also needs to include corrections to the latter
from the interference of one and two-photon exchange amplitudes
(denoted by ``$\gamma(\gamma\gamma)$'').
These have been computed in Ref.~\cite{BMT} within the current
framework.

The $\sw$ dependence of the TBE corrections can be obtained explicitly
by calculating separately the proton and neutron contributions of
Eq.~(\ref{eq:FZ}) to the PV asymmetry.
In so doing the $\sw$-dependent and independent parts can be evaluated
and the $\Delta\rho$ and $\Delta\kappa$ corrections determined from the
vector part of $A_{\rm PV}$ in Eq.~(\ref{eq:AV}):
\begin{subequations}
\label{eq:Drk}
\begin{align}
\label{eq:Dr}
\Delta\rho
&= { A_V^p + A_V^n \over A_V^{p, {\rm tree}} + A_V^{n, {\rm tree}} }
 - {\Delta\sigma^{\gamma(\gamma\gamma)} \over \sigma_{\rm red}}\ , \\
\label{eq:Dk}
\Delta\kappa
&= { A_V^p \over A_V^{p, {\rm tree}} }
 - { A_V^p + A_V^n \over A_V^{p, {\rm tree}} + A_V^{n, {\rm tree}} }\ ,
\end{align}
\end{subequations}
where $A_V^{p(n)}$ is the TBE contribution to $A_V$ from the proton
(neutron), and $A_V^{p(n), {\rm tree}}$ is the corresponding tree-level
asymmetry, with $\Delta\sigma^{\gamma(\gamma\gamma)}$ the electromagnetic
two-photon exchange contribution to $\sigma_{\rm red}$.
The contributions to $\Delta\rho$ arise therefore from the
$\gamma(Z\gamma)$ and $Z(\gamma\gamma)$ corrections, as well as
from the electromagnetic corrections $\gamma(\gamma\gamma)$.
On the other hand, $\Delta\kappa$ receives contributions only
from the $\gamma(Z\gamma)$ and $Z(\gamma\gamma)$ corrections.

The calculation of the TBE corrections proceeds along the same lines
as that of the $2\gamma$ amplitudes in Ref.~\cite{BMT}, with the
replacement of the $\gamma N N$ vertex function by $\Gamma_Z^\mu$
in Eq.~(\ref{eq:JZ}).
As in Ref.~\cite{BMT}, we parameterize the form factors $F_{1,2}^{Z p}$
as sums of three monopoles, but take the proton form factors from the
more recent global fit of Ref.~\cite{AMT}, and the neutron form factors
from Ref.~\cite{Bosted}.
Since the main purpose of the PV experiments is to extract strange
contributions to form factors by comparing the measured asymmetry
with the predicted zero-strangeness asymmetry, in all our numerical
simulations we set the strange form factors to zero,
$F_{1,2}^s = 0 = G_A^s$.
For the axial-vector form factor we use the empirical dipole fit,
$G_A(Q^2) = G_A(0)/(1+Q^2/\Lambda_A^2)^2$, where $G_A(0)=1.267$ is
the axial-vector charge, and the mass parameter $\Lambda_A = 1$~GeV.
Varying $\Lambda_A$ by 20\% does not affect the results significantly.

\begin{figure}
\includegraphics[height=5cm]{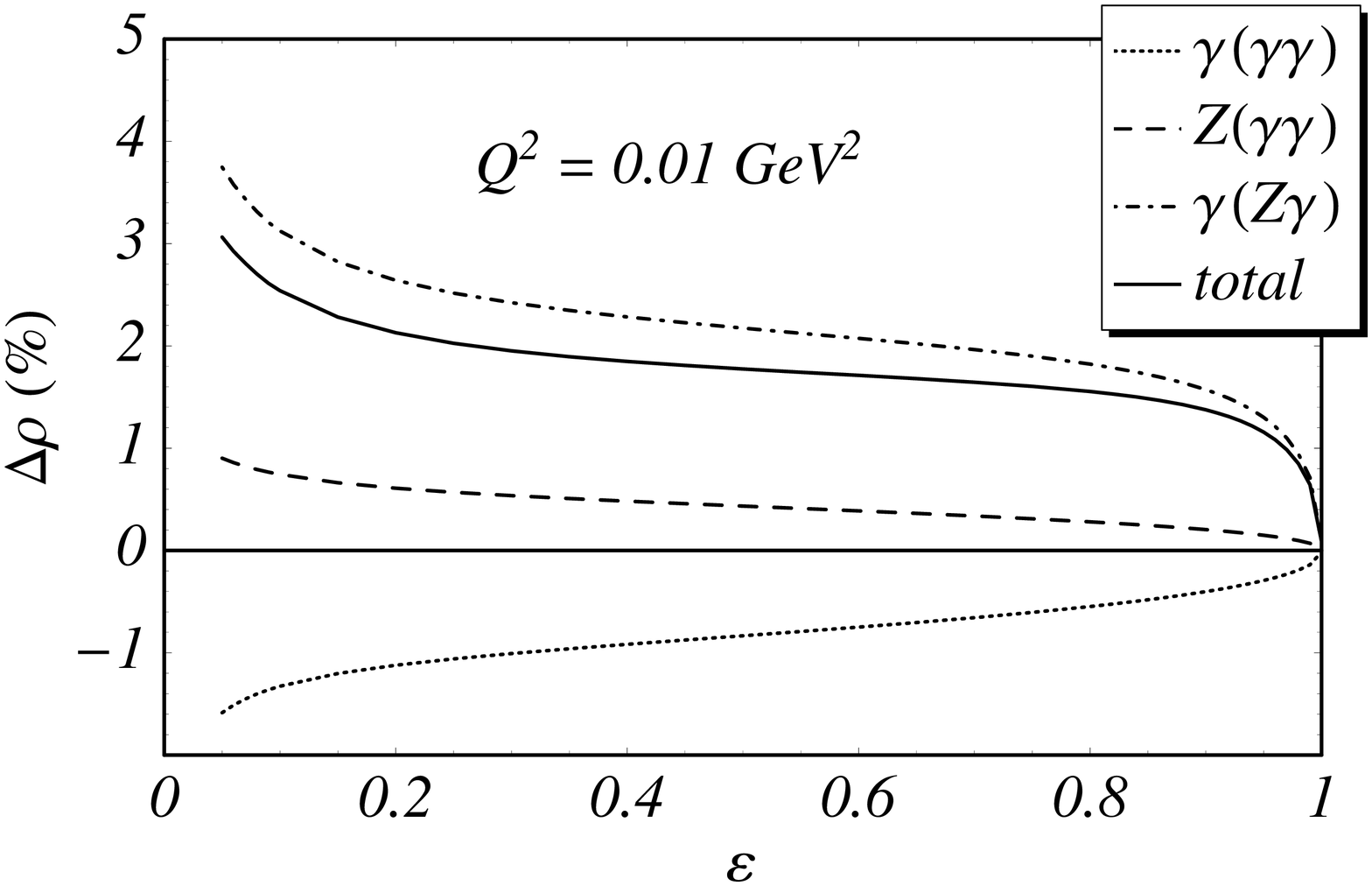}
\includegraphics[height=5cm]{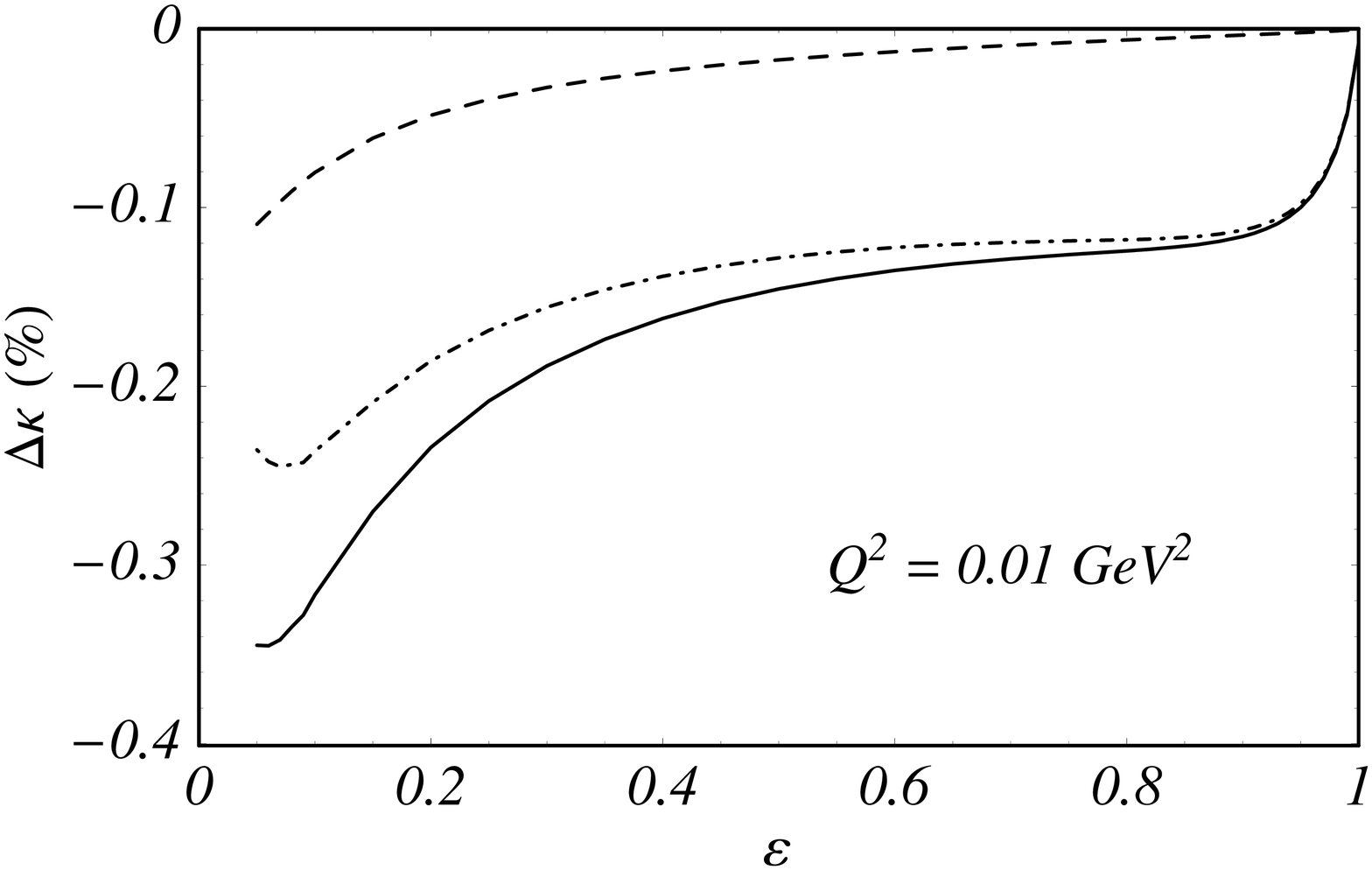}
\includegraphics[height=5cm]{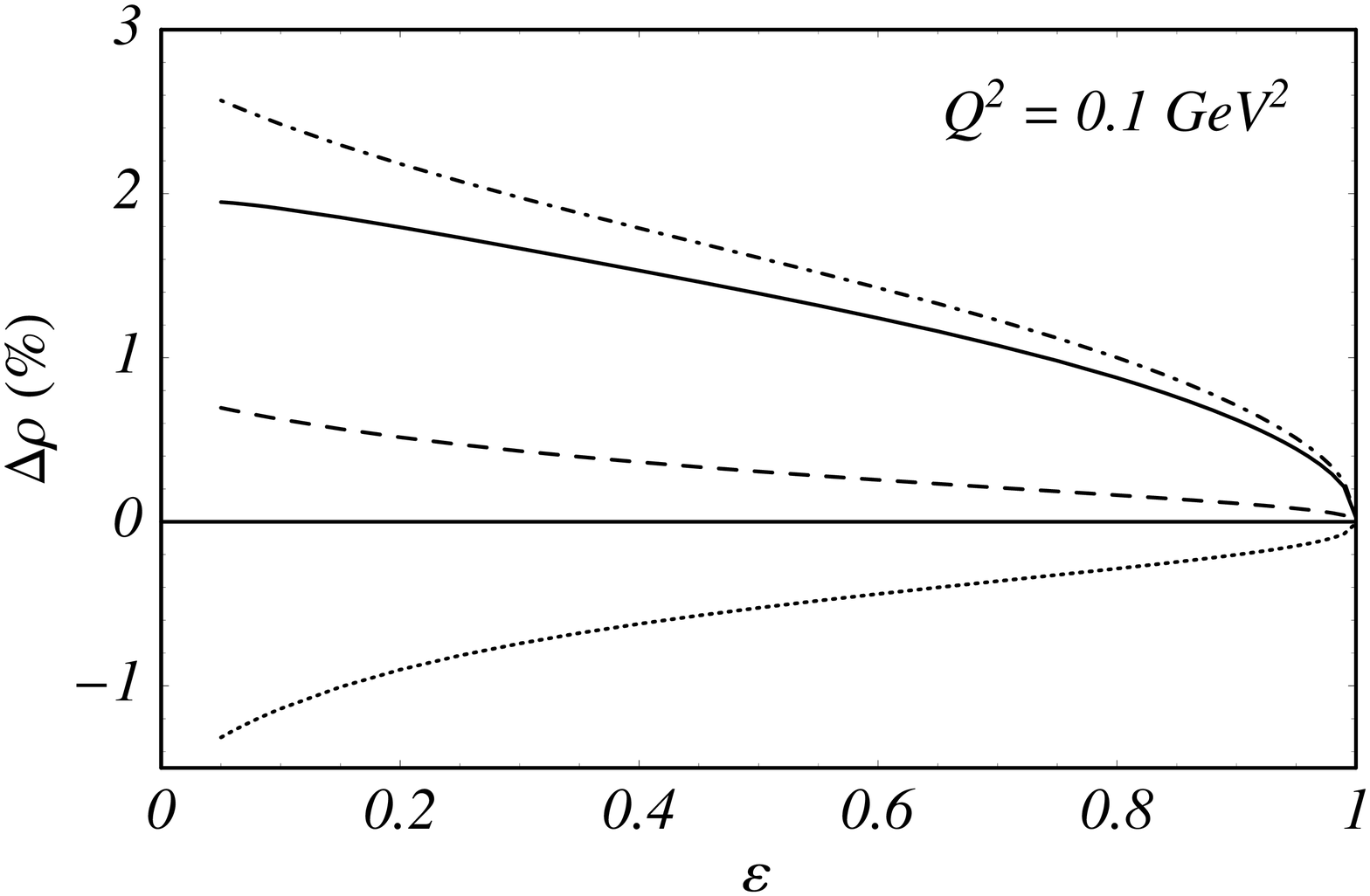}
\includegraphics[height=5cm]{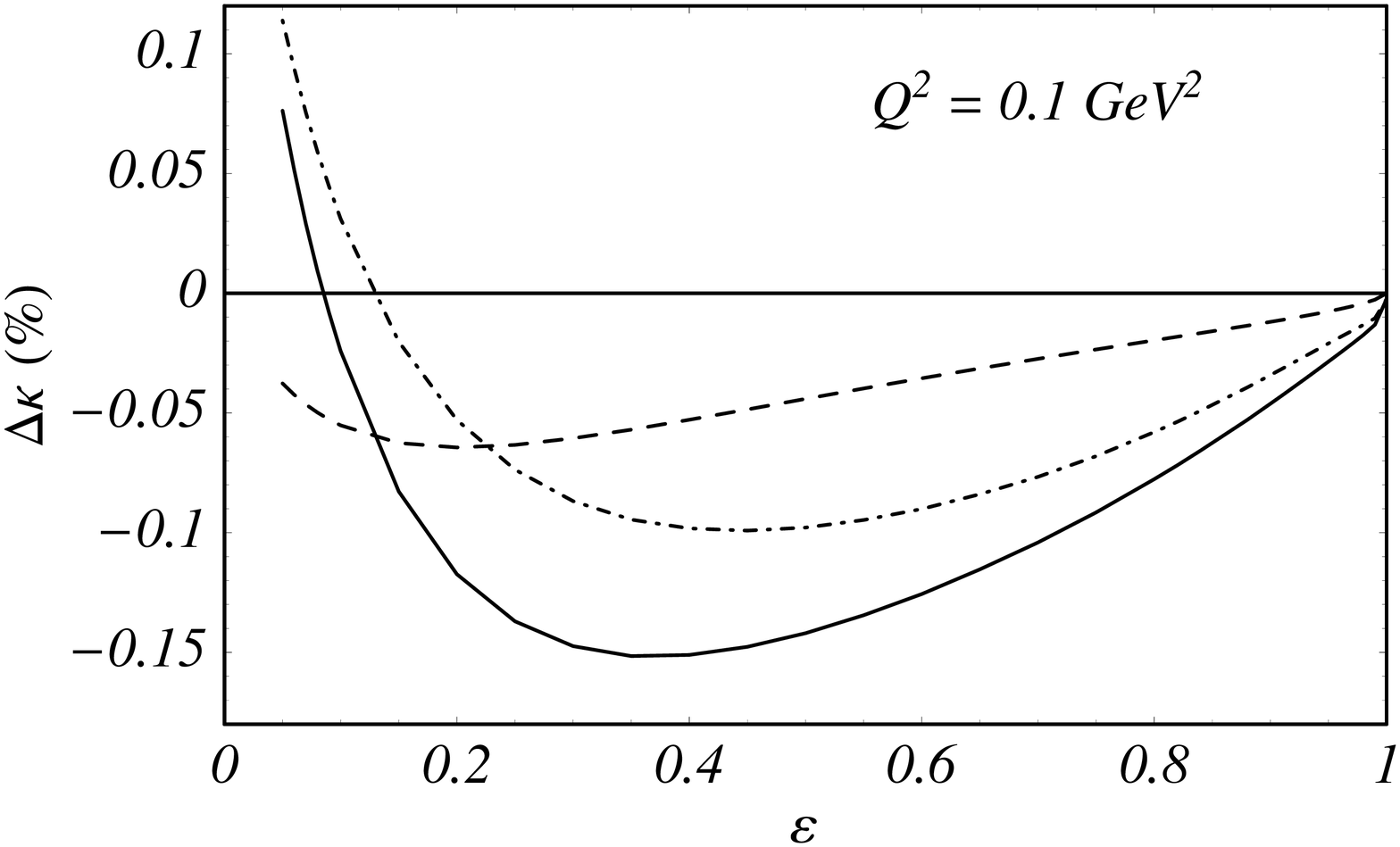}
\includegraphics[height=5cm]{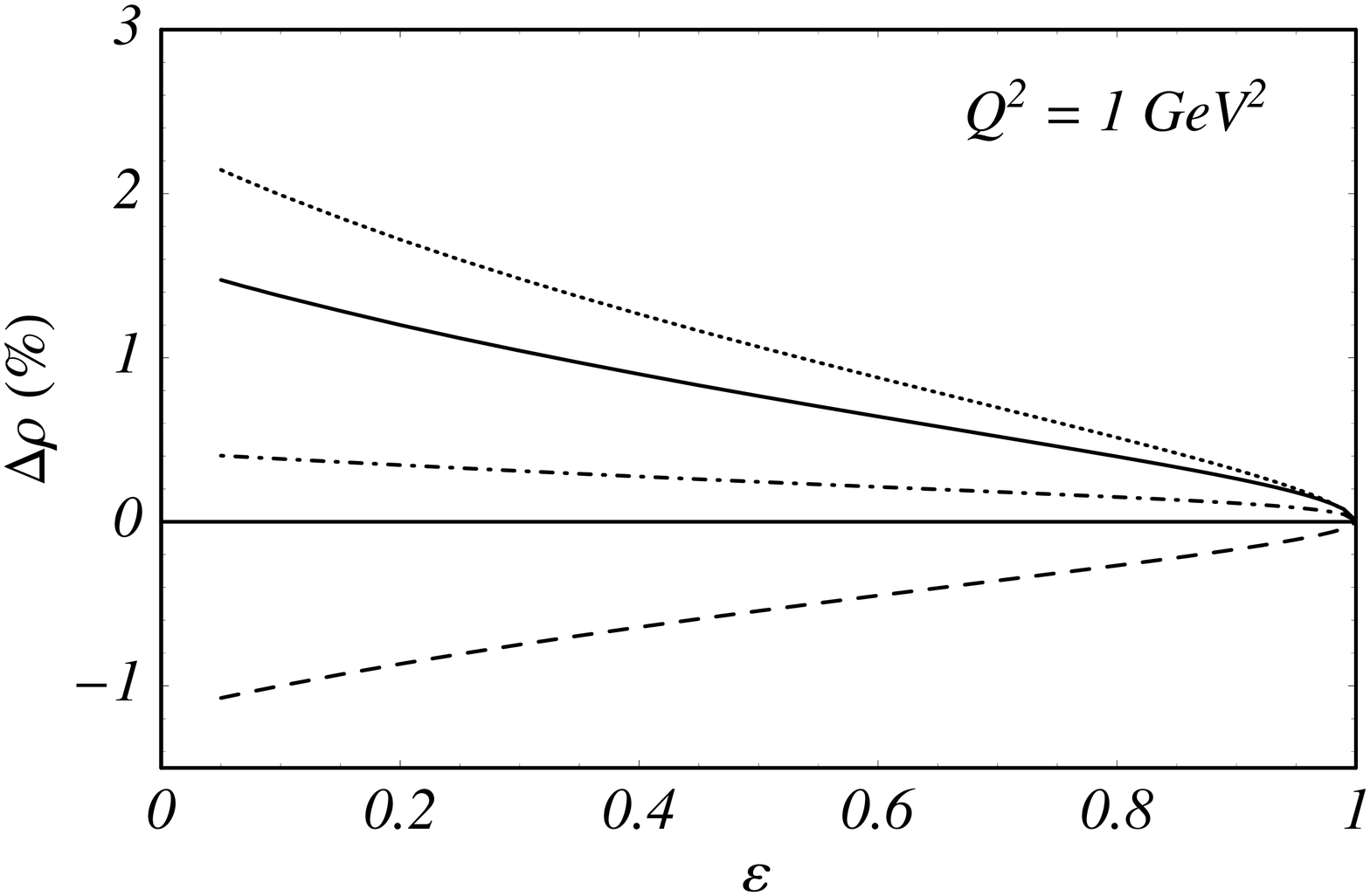}
\includegraphics[height=5cm]{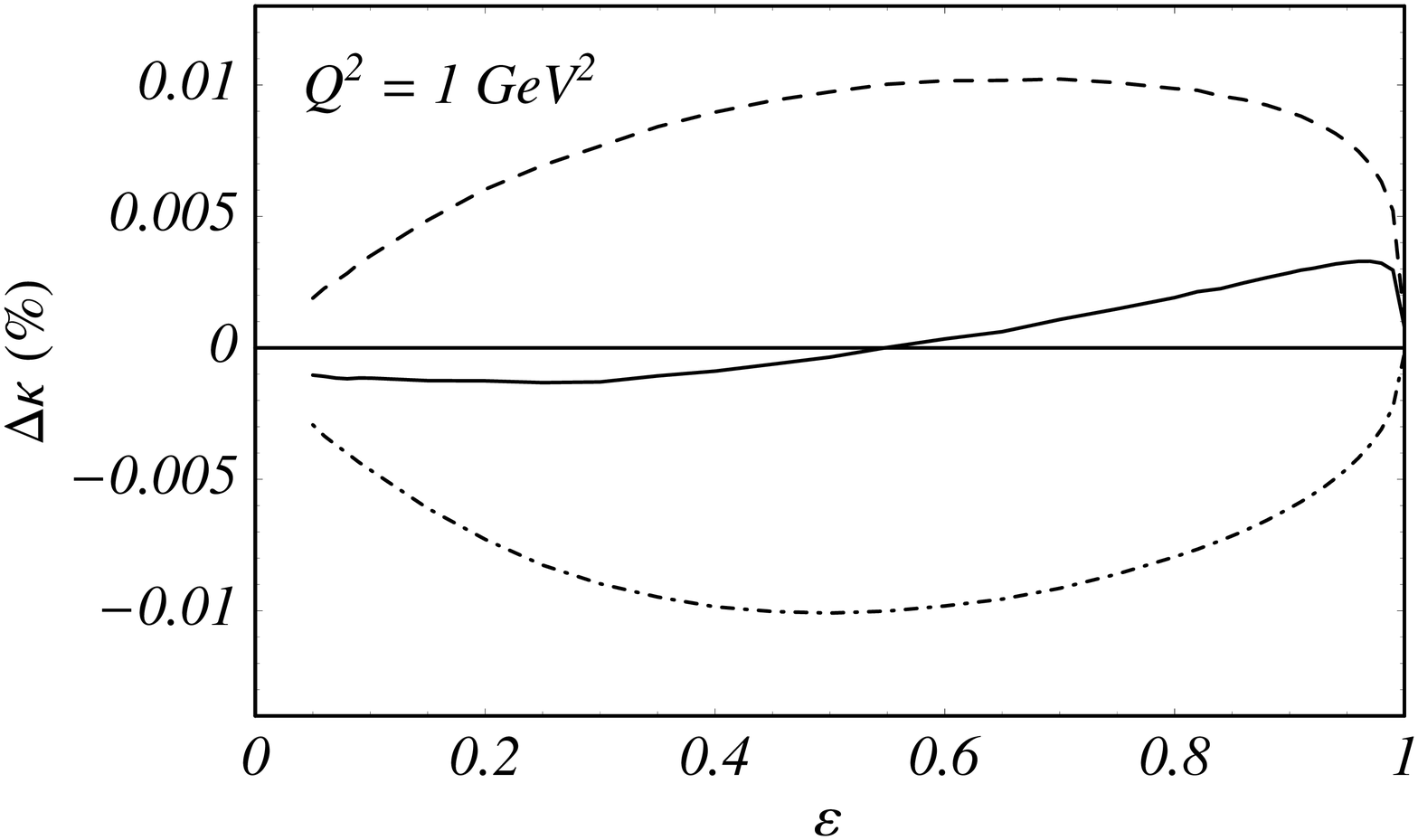}
\caption{TBE contributions $\Delta\rho$ and $\Delta\kappa$,
	relative to the SPA \cite{MT}, as a function of $\eps$
	for $Q^2 = 0.01, 0.1$ and 1~GeV$^2$.
	The various contributions are:
	$\gamma(\gamma\gamma)$ (dotted),
	$Z(\gamma\gamma)$ (dashed),
	$\gamma(Z\gamma)$ (dot-dashed),
	and the total (solid).}
\label{fig:rhokap}
\end{figure}

In Fig.~\ref{fig:rhokap} we show the calculated $\Delta\rho$ and
$\Delta\kappa$ corrections for $Q^2 = 0.01, 0.1$ and 1~GeV$^2$,
relative to the soft-photon approximation (SPA) of Mo \& Tsai \cite{MT}.
(All of the following results will be relative to the SPA.)
In the SPA the $Z$ exchange is factorized, which results in a
cancellation of the model independent infrared contribution to
the PV asymmetry.
The total $\Delta\rho$ is $\approx 1$--2\% over most of the range
of $\eps$, increasing at small $\eps$ and small $Q^2$.
At low $Q^2$ values the $Z(\gamma\gamma)$ piece largely cancels with
the $\gamma(\gamma\gamma)$, so that the total is saturated mostly by
the $\gamma(Z\gamma)$ contribution.
At $Q^2 = 1$~GeV$^2$ the signs of the $Z(\gamma\gamma)$ and
$\gamma(\gamma\gamma)$ corrections change, and the $\gamma(Z\gamma)$
contribution decreases in magnitude.

The TBE corrections to $\kappa$ are somewhat smaller in magnitude,
ranging from $\sim 0.2-0.3\%$ at $Q^2 = 0.01$~GeV$^2$, where the
$Z(\gamma\gamma)$ contribution is dominant, to less than 0.1\% at
$Q^2 = 1$~GeV$^2$, where there is large cancellation between these.
As seen in Eq.~(\ref{eq:Dk}), there is no contribution to
$\Delta\kappa$ from $\gamma(\gamma\gamma)$.

\begin{figure}
\includegraphics[height=5cm]{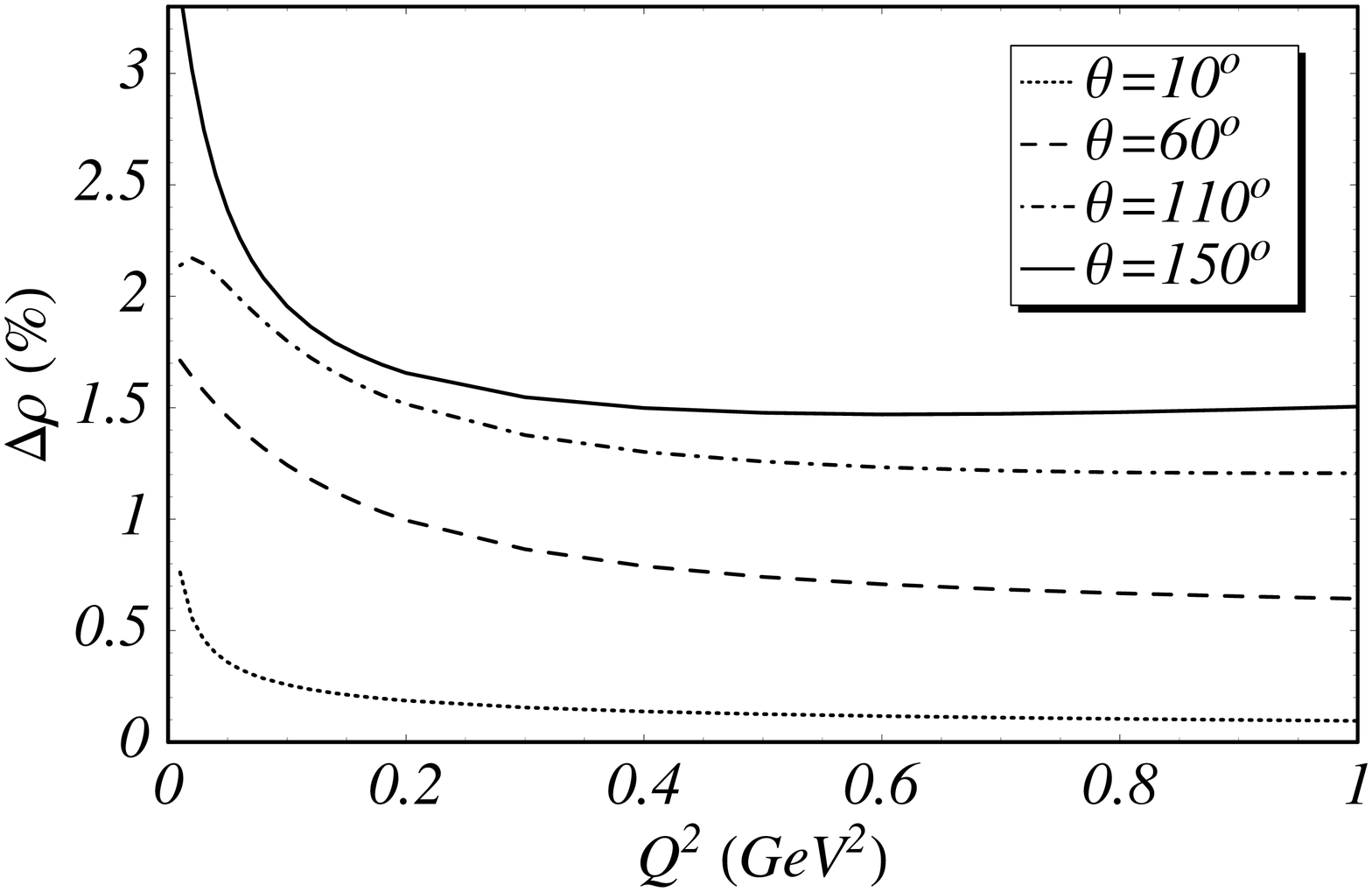}
\includegraphics[height=5cm]{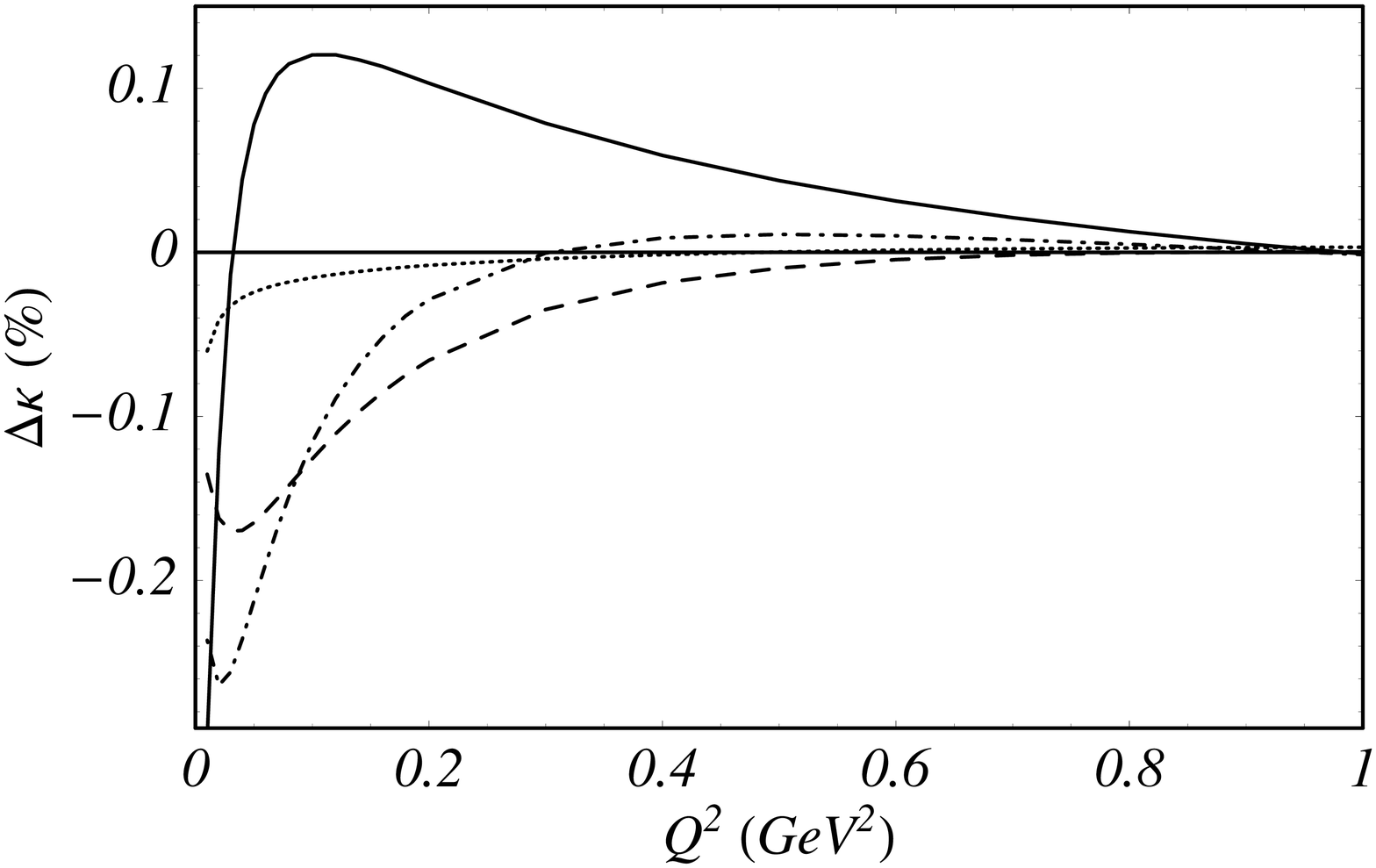}
\caption{TBE corrections $\Delta\rho$ and $\Delta\kappa$
	as a function of $Q^2$ for fixed scattering angle
	$\theta=10^\circ$ (dotted),
	$\theta=60^\circ$ (dashed),
	$\theta=110^\circ$ (dot-dashed), and
	$\theta=150^\circ$ (solid).}
\label{fig:ang}
\end{figure}

The $Q^2$ dependence of $\Delta\rho$ and $\Delta\kappa$ is
illustrated in Fig.~\ref{fig:ang} for several scattering angles.
A rapid $Q^2$ variation is evident for $Q^2 \lesssim 0.1$~GeV$^2$,
especially at backward scattering angles.
This is most pronounced in $\Delta\kappa$ and leads to a change
in sign at $Q^2 \approx 0.05$~GeV$^2$ at $\theta=150^\circ$.
Because of the strong $Q^2$ dependence, estimates at $Q^2=0$, by Marciano
and Sirlin \cite{Marciano} and more recently in Refs.~\cite{ABB,Erler}
in a somewhat different limit to the $Q^2=0$ point, are in general not
sufficient to obtain a reliable correction for the actual experiments.
Approximating the TBE corrections at non-zero $Q^2$ by their $Q^2=0$
values may therefore lead to errors in the extracted form factors.

\begin{table}
\caption{Ratio
	$\delta = A_{\rm PV}^{\rm TBE} / A_{\rm PV}^{\rm tree}$ of TBE
	contributions to the proton asymmetry relative to the tree-level
	asymmetry (in percent) at the $Q^2$ and scattering angle $\theta$
	of selected past and future experiments.
	The results are compared for different input form factors
	(empirical, dipole and monopole).}
\begin{tabular}{|ccc|ccc|}	\hline
$Q^2$~(GeV$^2$) & $\theta$ & Ref.
			&\multicolumn{3}{c|}{$\delta\ (\%)$} \\ \cline{4-6}
		&	&
			& empirical & dipole & monopole \\ \hline
0.1	& 144.0$^\circ$	& \cite{SAMPLE97}
			& 1.62	& 1.52	& 1.72		\\
0.23	& 35.31$^\circ$	& \cite{PVA404}
			& 0.63	& 0.58	& 0.84		\\
0.477	& 12.3$^\circ$	& \cite{HAPPEX04}
			& 0.16	& 0.15 	& 0.24		\\
% 0.122	& 6.68$^\circ$	& \cite{G0}
%			& 0.19	& 0.16	& 0.31		\\
0.997	& 20.9$^\circ$	& \cite{G0}
			& 0.22	& 0.23	& 0.30		\\
0.109	& 6.0$^\circ$	& \cite{HAPPEX07}
			& 0.20	& 0.16	& 0.32		\\
0.23	& 110.0$^\circ$	& \cite{G0BACK}
			& 1.39	& 1.33	& 1.52		\\
0.03	& 8.0$^\circ$	& \cite{QWEAK}
			& 0.58	& 0.47	& 0.86		\\ \hline
\end{tabular}
\end{table}

The above behavior is qualitatively reproduced if one uses dipole
form factors (for either the Dirac and Pauli, or electric and magnetic
form factors, with a dipole mass of 1~GeV) or monopoles (for the
electric and magnetic, with a monopole mass of 0.56~GeV \cite{Yang}),
instead of the empirical ones \cite{AMT}.
Quantitatively, however, there are significant differences between
the empirical and monopole results at large $\eps$, with the monopole
results for $\Delta\rho$ ($\Delta\kappa$) being $\sim 30\%$ (60\%)
larger in magnitude at $\eps \sim 0.9$, with larger differences at
larger $Q^2$.
This reflects the sensitivity of the loop integrals to the
large-momentum tails of form factors (and hence nucleon structure
effects), even at low $Q^2$.

The sensitivity to the input form factors is more clearly illustrated
in Table~I, where we list the values of the TBE contributions to the
proton PV asymmetry relative to the tree-level asymmetry (in percent),
$\delta = A_{\rm PV}^{\rm TBE} / A_{\rm PV}^{\rm tree}$, at the
kinematics relevant to several past and future experiments
\cite{SAMPLE97,PVA404,HAPPEX04,G0,HAPPEX07,G0BACK,QWEAK}.
The results for the empirical form factors \cite{AMT} are compared with
those using dipole and monopole form factors in the loop integration.
Generally the effects using the empirical form factors are
$\lesssim 0.5\%$ for most of the forward angle experiments,
increasing to $\sim 1.5\%$ at backward angles.
The results with the dipole form factors are similar, tending to be
$\sim 5$--10\% smaller.
With the monopole form factors \cite{Yang}, however, the values of
$\delta$ are some 50\% larger than with the empirical, which suggests
insufficient suppression of contributions from large loop momenta.
The TBE corrections to $A_{\rm PV}$ may therefore be overestimated
using monopole form factors.

The impact of these differences on the strange form factors is
difficult to gauge without performing a full reanalysis of the data,
since in general different electroweak parameters and form factors
are used in the various experiments
\cite{SAMPLE97,PVA404,HAPPEX04,G0,HAPPEX07,G0BACK,QWEAK}.
However, as an estimate of the possible effects we have determined
following Ref.~\cite{Yang} the quantity defined there as $\delta_G$,
as a measure of the induced difference between the strange asymmetry
extracted using the different form factors.
With the parameters of this analysis we find, for example, differences
of the order of 15\% between the empirical and monopole form factors
for the HAPPEX kinematics \cite{HAPPEX04}, around 20\% for the G0
datum \cite{G0} in Table~I, and over 30\% for the PVA4 kinematics
\cite{PVA404}.
Although these values should be treated as indicative only, they 
clearly point to the need for a more detailed reanalysis of the data
including TBE effects and a careful treatment of the form factors in
the loop integrations.

\begin{figure}
\includegraphics[height=6cm]{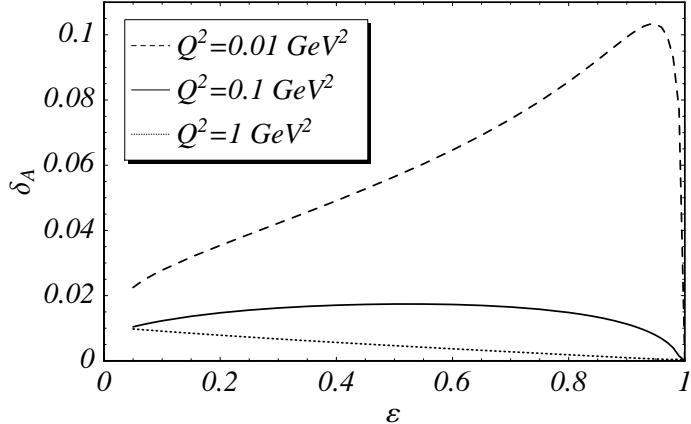}
\caption{TBE correction $\delta_A$ to the hadronic axial part
	of the PV asymmetry, for $Q^2 = 0.01$ (dashed),
	0.1 (solid) and 1~GeV$^2$ (dotted).}
\label{fig:dA}
\end{figure}

As discussed above, the effective axial-vector form factor
$\widetilde{G}_A^{Z p}$ in the axial asymmetry $A_A$ is defined
to include the anapole form factor, and higher order radiative
corrections.
In extracting the strange form factors from data, Young {\em et al.}
\cite{Ross} fit the effective $\widetilde{G}_A^{Z p}$ without
decomposing it into its various contributions.
Alternatively, one may extract the anapole form factor from the
$\widetilde{G}_A^{Z p}$ by correcting for the radiative effects.
In Fig.~\ref{fig:dA} we show the correction $\delta_A$ to the
axial PV asymmetry, $A_A \to A_A (1 + \delta_A)$, at several $Q^2$
values.
The correction is of order 1--2\% for $Q^2 \geq 0.1$~GeV$^2$, but
increases rapidly for decreasing $Q^2$, especially at large $\eps$,
where it reaches 10\% at $\eps \sim 0.95$ at $Q^2 = 0.01$~GeV$^2$.
This behavior may be related to the faster vanishing of the axial
Born asymmetry compared with the TBE contribution at large $\eps$.

The corrections calculated here are presented in a form that can
be straightforwardly applied to the $A_{\rm PV}$ data.
The values for $\Delta\rho$ and $\Delta\kappa$ can simply be
added to the existing radiative corrections contained in $\rho$
and $\kappa$, taking care to subtract any partial TBE contributions
that have already been included \cite{Marciano}.
Because the TBE effects are largest at backward angles, they will
be most relevant for the SAMPLE experiment at Bates \cite{SAMPLE97},
and for the backward angle run of the G0 experiment \cite{G0} at
Jefferson Lab.
For the former, we find
$(\Delta\rho, \Delta\kappa)(Q^2=0.1~{\rm GeV}^2,\theta=144^\circ)
= (1.950 \times 10^{-2}, 7.998 \times 10^{-4})$,
while for the latter
$(\Delta\rho, \Delta\kappa)(Q^2=0.23~{\rm GeV}^2,\theta=110^\circ)
= (1.473 \times 10^{-2}, -1.342 \times 10^{-4})$
and
$(\Delta\rho, \Delta\kappa)(Q^2=0.62~{\rm GeV}^2,\theta=110^\circ)
= (1.261 \times 10^{-2}, 1.103 \times 10^{-4})$.
In addition, even though it is at forward angles, the considerably
smaller uncertainties expected in the Qweak experiment \cite{QWEAK}
at Jefferson Lab will require a careful treatment of the radiative
effects.
In particular, we find
$(\Delta\rho, \Delta\kappa)(Q^2=0.03~{\rm GeV}^2,\theta=8^\circ)
= (3.755 \times 10^{-3}, -2.655 \times 10^{-4})$.
A reanalysis of the entire data set on strange form factors
incorporating these effects is currently in progress.

%%%%%%%%%%%%%%%%%%%%%%%%%%%%%%%%%%%%%%%%%%%%%%%%%%%%%%%%%%%%%%%%%%%%%%%%%
\begin{acknowledgments}

We thank A.~W.~Thomas and R.~D.~Young for helpful comments and
suggestions.
W.~M. is supported by the DOE contract No. DE-AC05-06OR23177, under
which Jefferson Science Associates, LLC operates Jefferson Lab.

\end{acknowledgments}

%%%%%%%%%%%%%%%%%%%%%%%%%%%%%%%%%%%%%%%%%%%%%%%%%%%%%%%%%%%%%%%%%%%%%%%%%

\end{document}